\documentclass[aps,pra,twocolumn,superscriptaddress,longbibliography]{revtex4-1}

\usepackage{amsmath}
\usepackage{amssymb}
\usepackage{graphicx}
\usepackage{dsfont}
\usepackage[colorlinks = true, linkcolor = blue, citecolor = blue,urlcolor = blue]{hyperref}
\usepackage{natbib}

\usepackage[utf8]{inputenc}
\usepackage{blindtext}
\usepackage{enumerate}
\usepackage[caption=false]{subfig}
\renewcommand{\vec}[1]{\boldsymbol{#1}}

\newcommand{\oH}{\hat{ \mathcal H}}

\newcommand{\vS}{\vec{S}}
\newcommand{\oS}{\hat{S}}
\newcommand{\ovS}{\vec{\hat{S}}}

\newcommand{\oa}{\hat{a}}
\newcommand{\oad}{\hat{a}^\dagger}

\newcommand{\oal}{\hat{\alpha}}
\newcommand{\oald}{\hat{\alpha}^\dagger}

\newcommand{\muB}{\mu_{\textrm{B}} g B_{\boldsymbol{n}}}

\newcommand{\on}[1]{\hat{a}^\dagger_{#1}\hat{a}^{\phantom{\dagger}}_{#1} }
\newcommand{\onal}[1]{\hat{\alpha}^\dagger_{#1}\hat{\alpha}^{\phantom{\dagger}}_{#1} }

\begin{document}

\title{Magnon squeezing in conical spin spirals}
\author{D. Wuhrer}
\email{dennis.wuhrer@uni-konstanz.de}
\affiliation{Fachbereich  Physik,  Universit\"at  Konstanz,  D-78457  Konstanz,  Germany}
\author{L. Rózsa}
\email{levente.rozsa@uni-konstanz.de}
\affiliation{Fachbereich  Physik,  Universit\"at  Konstanz,  D-78457  Konstanz,  Germany}
\affiliation{Department of Theoretical Solid State Physics, Institute of Solid State Physics and Optics, Wigner Research Centre for Physics, H-1525 Budapest, Hungary}
\affiliation{Department of Theoretical Physics, Budapest University of Technology and Economics, H-1111 Budapest, Hungary}
\author{U. Nowak}
\affiliation{Fachbereich  Physik,  Universit\"at  Konstanz,  D-78457  Konstanz,  Germany}
\author{W. Belzig}
\affiliation{Fachbereich  Physik,  Universit\"at  Konstanz,  D-78457  Konstanz,  Germany}

\begin{abstract}
	\noindent  
	We investigate squeezing of magnons in a conical spin spiral configuration. We find that while the energy of magnons propagating along the $\vec{k}$ and the $-\vec{k}$ directions can be different due to the non-reciprocal dispersion, these two modes are connected by the squeezing, hence can be described by the same squeezing parameter. The squeezing parameter diverges at the center of the Brillouin zone due to the translational Goldstone mode of the system, but the squeezing also vanishes for certain wave vectors. We discuss possible ways of detecting the squeezing. 
\end{abstract}

{\hypersetup{urlcolor = black}
\maketitle
}

\section{Introduction}
Magnons are collective excitations of magnetically ordered systems, which may be interpreted as a wave propagating through the material carrying spin angular momentum and magnetic moment~\cite{Spin_Waves_Dyson,Spin_Waves_Kittel,Spin_Waves_NoltingRamakanth}. Due to their relatively low dissipation, magnons have risen as possible candidates to process and transport information in computing architectures~\cite{Chumak2015,Magnonics_Yuan,Magnonics_Squeezing_Akash, Barman_2021}.

Recently there is a growing interest in taking advantage of the quantum-mechanical nature of magnons~\cite{Magnonics_Yuan}. One of these properties is squeezing, where the uncertainty in one observable is reduced at the cost of an increased variance in the conjugated observable~\cite{Magnonics_Squeezing_Akash}. The squeezing implies entanglement both in the ground state of the system~\cite{Non_Integer_Spin_Akash,AFM_Squeezing_Akash, AFM_Squeezing_Wuhrer, Zou_Tserkovnyak_1} as well as in the excited states carrying a non-integer multiple of $\hbar$ as spin~\cite{Non_Integer_Spin_Akash}. This entanglement may be utilized by coupling the magnetic system to quantum dots~\cite{Multiple_QDot_Excitation_Skogvoll, Zou_Tserkovnyak_2}.

In contrast to squeezing in photonic systems which may be achieved under non-equilibrium conditions~\cite{SqueezedLight_Walls, SqueezedLight_Wu, SqueezedLight_knight}, the squeezing of magnons is an intrinsic property. The degree of squeezing in ferromagnets can already achieve relatively large values compared to photonic systems~\cite{Non_Integer_Spin_Akash}, where it is caused by the relatively weak magnetocrystalline and dipolar anisotropy terms. Squeezing is further enhanced in antiferromagnets~\cite{AFM_Squeezing_Akash,AFM_Squeezing_Wuhrer}, where the Heisenberg exchange interaction is responsible for the squeezing, which is typically the largest magnetic energy scale. Magnon squeezing has also been studied in two-sublattice ferrimagnets~\cite{Ferrimagnet_Squeezing_Akash}, which interpolate between the ferromagnetic and antiferromagnetic limits by tuning the magnitude of the magnetic moment on one of the sublattices.

An alternative approach to transforming the parallel spin alignment in ferromagnets to the antiparallel alignment in antiferromagnets is by continuously increasing the angle between neighboring spins, leading to the formation of a spin spiral. Spin spirals may be stabilized by the competition between ferromagnetic and antiferromagnetic exchange interactions with different neighbors, as is common in, e.g., the rare-earth metals Ho, Tb or Dy~\cite{Izyumov_1984}. The Dzyaloshinsky--Moriya interaction (DMI)~\cite{DMI_Dzyaloshinskii,DMI_Moriya} is present in materials with broken inversion symmetry, where it creates spin spirals with a preferred rotational sense. These have been studied extensively over the last decades both in bulk samples, such as the B20 class including FeCoSi, MnSi or Cu$_{2}$OSeO$_{3}$~\cite{Uchida2006,Seki2012,Bauer2012}, as well as in atomically thin magnetic layers including Mn mono- and double-layers on W(110)~\cite{Bode2007,Yoshida2012,VonBergmann2014,Hasselberg2015}. A planar spin spiral state often transforms into a conical spin spiral under the application of an external magnetic field, possessing a finite net magnetic moment along the cone axis parallel to the field. The external field may also be utilized to create magnetic domain walls or skyrmions~\cite{Bogdanov1989,Nagaosa2013}, which themselves have been suggested as suitable candidates for unconventional information processing~\cite{Racetrack_Memory_Hayashi, Racetrack_Memory_Parkin,Fert2013}. Magnon excitations of spin spirals and skyrmions have been analyzed in the classical limit~\cite{Garst2017}, for example from the standpoint of the topology of the magnon band structure~\cite{Weber2022}. Quantum effects and in particular squeezing in such magnetic configurations seem to have eluded attention so far.

In this work, we investigate magnon squeezing in conical spin spirals stabilized by frustrated Heisenberg interactions and DMI in the presence of an external field. The system enables the analytical description of the magnon dispersion~\cite{SpinSpiralDispersion_Michael}, providing a clear insight into the entanglement between the modes. The considered system displays non-reciprocal spin-wave propagation common in non-collinear spin structures~\cite{Garst2017}, meaning that the frequency of magnons with opposite wave vectors differs from each other. We find a high degree of squeezing which typically increases with the angle between the spins when going from the ferromagnetic toward the antiferromagnetic configuration. The squeezing parameter typically decreases when moving away from the center of the Brillouin zone, but we find certain curves along which it exactly vanishes. We establish that the non-reciprocity of the magnon dispersion is not observed in the squeezing parameter due to the entanglement between the $\boldsymbol{k}$ and $-\boldsymbol{k}$ modes.

This work is structured as follows. In Sec.~\ref{sec:2} we discuss the theory of squeezing in a general spin model forming a conical spin spiral ground state in the classical limit.
We discuss in Sec.~\ref{subsec:2a} the properties of the ground state. 
Sec.~\ref{subsec:2b} copes with the determination of the magnon dispersion. 
Subsequently, in Sec.~\ref{subsec:2c}, we calculate the squeezing parameter $r_{\vec k}$ to quantify the degree of squeezing. 
Finally, in Sec.~\ref{sec:3}, as a specific example we discuss a two-dimensional square lattice magnet on a substrate with nearest-neighbor (NN) and next-nearest-neighbor (NNN) Heisenberg interaction and NN DMI.

\section{General spin spirals}
\label{sec:2}
We consider the Hamiltonian
\begin{align}
    \oH = \oH_{\text{H}} + \oH_{\text{DMI}}  +\oH_{\text{Z}}\, ,\label{eq:Def_Ham}
\end{align}
where 
\begin{align}
    \oH_{\text{ex}} = \frac 1 2 \sum_{\vec R_i,\vec R_j} J_{ij} \ovS_i \cdot \ovS_j\, ,
    \label{eq:Heisenberg}
\end{align}
with $J_{ij}$ being the strength of the symmetric Heisenberg exchange interaction between spins at sites $\vec R_i$ and $\vec R_j$. Note that with this sign convention, negative and positive values of $J_{ij}$ denote ferromagnetic and antiferromagnetic coupling, respectively. $\ovS_i$ stands for the dimensionless spin operator at site $\vec{R}_{i}$. 
The DMI is given by
\begin{align}
\begin{split}
    \oH_{\text{DMI}} &= \frac 1 2 \sum_{\vec R_i,\vec R_j} \vec D_{ij} \cdot \left(\ovS_i \times \ovS_j\right)\, .
\label{eq:DMI_Def}
\end{split}
\end{align}
$\vec D_{ij}$ expresses the strength of the antisymmetric exchange between spins at sites $\vec R_i$ and $\vec R_j$. It is a vector quantity whose direction strongly depends on the lattice symmetry as shown by Moriya \cite{DMI_Moriya}. As an antisymmetric interaction, the DMI changes sign under exchange of the positions $\vec D_{ij}=-\vec{D}_{ji}$. 

The last term in Eq.~\eqref{eq:Def_Ham} reads 
\begin{align}
    \oH_{\text{Z}} &= -\muB \sum_{\vec R_i} \vec{n}\ovS_{i} \,,
    \label{eq:Anisotropy_Zeeman}
\end{align}
with $B_{\vec{n}}$ being 
the external magnetic field oriented along the unit vector $\vec{n}$ in this model. 

\subsection{Classical ground state}
\label{subsec:2a}
\begin{figure}
	\begin{center}
			\includegraphics[width = \columnwidth]{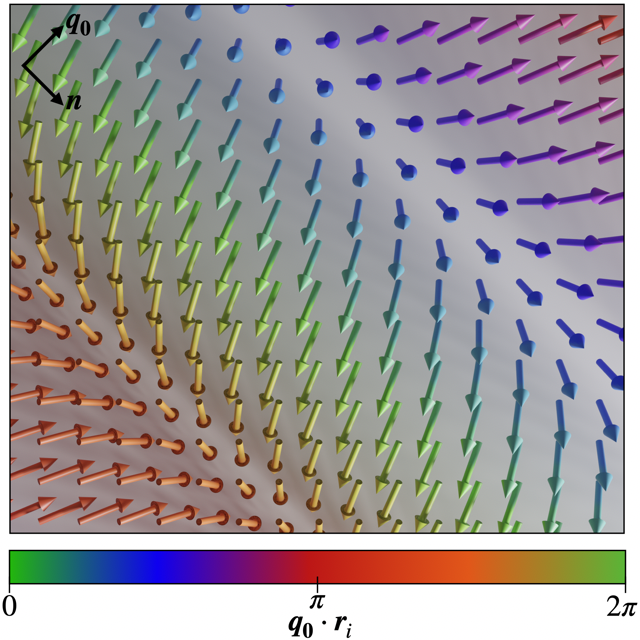}
	\end{center}
	\caption{Illustration of the classical conical spin spiral state. The configuration is the ground state of the two-dimensional square lattice model discussed in Sec.~\ref{sec:3}, with $J_2 = D = 0.2  |J_1|$, $\muB = 0.025  |J_1|$ and $\vartheta \approx 0.4 \pi$. The color indicates the phase of the rotation $\vec r_i \cdot \vec q_0$ around the cone opening direction $\vec n$.
	}
	\label{fig:1}
\end{figure}

As a starting point for the quantisation procedure, we determine the classical spin configuration minimizing the  
energy. For this we substitute $\ovS_i$ by $\vS_i = \left<\ovS_i\right>$ in the Hamiltonian, with the ansatz 
\begin{align}
    \vS_i = S&\left[\sin\left(\vec R_i \cdot\vec q_0\right)\sin (\vartheta)\vec{e}_{1}+\cos\left(\vec R_i \cdot\vec q_0\right)\sin (\vartheta)\vec{e}_{2}\right.\nonumber\\
    &\left.+\cos (\vartheta)\vec{n}\right],
\end{align}
where the unit vectors $\vec{e}_{1},\vec{e}_{2},$ and $\vec{n}$ form a right-handed system. This expression describes a harmonic conical spin spiral state with the axis of the cone along the $\vec{n}$ direction, as illustrated in Fig.~\ref{fig:1}. Here $S$ is the spin quantum number, $\vartheta$ is the opening angle of the cone and $\vec q_0$ is the wave vector of the spin spiral. In Eq.~\eqref{eq:Anisotropy_Zeeman} it was assumed that the magnetic field is pointing along the cone axis $\vec{n}$. Deviations from this would induce distortions in the spin spiral compared to the harmonic form, making a general analytical treatment impossible. 

Substituting $\vS_i$ into Eq.~\eqref{eq:Def_Ham} we end up with the classical energy
\begin{align}
    \begin{split}
        E_{\textrm{cl}} =&   \frac{S N}{2}\left[ S \left(\cos^2(\vartheta) J_0 + \sin^2(\vartheta)\left( J_{\vec q_0} + iD_{\vec q_0}^{\vec{n}}\right)\right)\right. -\\
         &\left.  \phantom{\frac{NS}2}- 2 \muB \cos (\vartheta)\right] \, . \label{eq:Classical_Energy}
    \end{split}
\end{align}
$J_{\vec q_0}$ and $D_{\vec q_0}^{\vec{n}}$ are the spatial Fourier transforms of the symmetric and antisymmetric exchange interactions for wave vector $\vec q_0$, 
\begin{align}
    J_{\vec{q}_{0}} &= \sum_{\vec \delta} J_{\vec \delta} e^{-i \vec \delta \cdot \vec{q}_{0}}  \, ,&
     \vec D_{\vec{q}_{0}}= \sum_{\vec \delta}  \vec D_{\vec \delta} e^{-i \vec \delta \cdot \vec{q}_{0}} \, ,
\end{align}
where $\vec \delta=\vec{R}_{i}-\vec{R}_{j}$ runs over all possible lattice vectors. 
$D_{\vec q_0}^{\vec{n}} = \vec n \cdot \vec D_{\vec q_0}$ is given by the projection of the Fourier transform of the DMI onto the cone opening direction. 

The exact form of Fourier transforms of the DMI and the Heisenberg interaction depend on the symmetry of the considered lattice chosen. One should note that since $J_{ij}$ is symmetric, $J_{-\vec q_0} = J_{\vec q_0}$ is always a real symmetric quantity, while the antisymmetry of $\vec{D}_{ij}$ implies $\vec{D}_{-\vec q_0}^{\vec{n}} = -\vec{D}_{\vec q_0}^{\vec{n}}$ to be purely imaginary, i.e., $i \vec{D}_{\vec q_0}^{\vec{n}}$ is real and antisymmetric. 

We need to minimize Eq.~\eqref{eq:Classical_Energy} with respect to the 
parameters $\vartheta$ and $\vec q_0$. Minimizing the energy with respect to the opening angle $\vartheta$ of the spin cone yields two possible solutions,
\begin{align}
    \sin \vartheta = 0 \quad\text{and}\quad \cos{\vartheta}=   \frac{\muB}{A_{\vec q_0}}\, ,\label{eq:Theta_min}
\end{align}
where the latter depends on $A_{\vec q_0}= S\left[ J_{\vec 0} - \left(J_{\vec q_0} + i D^{\vec{n}}_{\vec q_0} \right) \right]$ and therefore on $\vec q_0$. The minimum is given by $\sin \vartheta = 0$ for $A_{\vec q_0} < 0$ or $|\muB| > |A_{\vec q_0}|$. In this case, the spins align ferromagnetically along the magnetic field direction. The second solution describes a spin spiral ground state, which is planar ($\vartheta=\pi/2$) for $B_{{\vec{n}}}=0$ and conical otherwise. Increasing the magnetic field eventually closes the cone and forces the system into the collinear configuration. 

Minimizing the energy with respect to $\vec q_0$ is equivalent to solving 
\begin{align}
    \partial_{\vec q_0} \left(J_{\vec q_0} + i D_{\vec q_0}^{\vec{n}}\right) &= 0\, , \label{eq:q0_min}
\end{align}
and proving that the solution is a minimum. This condition implicitly depends on the choice of the ${\vec{n}}$ direction through $D_{\vec q_0}^{\vec{n}}$. We use the following procedure for the minimization: First, we determine $\vec D_{\vec q_0}$ based on the symmetries of the system~\cite{DMI_Moriya}. Second, for each $\vec{q}_0$ we set the ${\vec{n}}$ direction to be antiparallel to $i\vec D_{\vec q_0}$, since this minimizes $iD_{\vec q_0}^{\vec{n}}$ when $\vec{q}_{0}$ is fixed. Third, we solve Eq.~\eqref{eq:q0_min} to find the wave vector. Note that this establishes a connection between the direction of $\vec{q}_{0}$ and the ${\vec{n}}$ direction. Since Eq.~\eqref{eq:q0_min} is independent of the magnetic field, this only means that the field must be oriented along the ${\vec{n}}$ direction determined from the minimization procedure. 
Fourth, we calculate $\vartheta$ from Eq.~\eqref{eq:Theta_min}. Note that the wave vector specified by Eq.~\eqref{eq:q0_min} does not depend on the opening angle $\vartheta$ assuming that the configuration is a real spin spiral, i.e., $\sin\vartheta\neq 0$.

An example for determining the classical ground state will be discussed for a specific system in Sec.~\ref{sec:3}.

\subsection{Magnon spectrum}
\label{subsec:2b}
We determine the single-particle excitations of the quantum-mechanical system Eq.~\eqref{eq:Def_Ham} by an expansion around the classical ground state. With the parameters given by Eqs.~\eqref{eq:Theta_min} and \eqref{eq:q0_min} we define 
\begin{gather}
    \vec e_{i,1} =  \begin{pmatrix}
        \sin\left(\vec R_i \cdot\vec q_0\right)\cos (\vartheta)\\\cos\left(\vec R_i \cdot\vec q_0\right)\cos (\vartheta)\\ - \sin (\vartheta)
    \end{pmatrix}\, ,  \vec e_{i,2} =  \begin{pmatrix}
    -\cos\left(\vec R_i \cdot\vec q_0\right)\\  \sin\left(\vec R_i \cdot\vec q_0\right)\\ 0
    \end{pmatrix}\, ,\notag\\  \vec e_{i,3} = \frac{1}{S}\vS_i = \begin{pmatrix}
    \sin\left(\vec R_i \cdot\vec q_0\right)\sin (\vartheta)\\\cos\left(\vec R_i \cdot\vec q_0\right)\sin (\vartheta)\\\cos (\vartheta)
    \end{pmatrix}\, ,
    \label{eq:eigenvectors}
\end{gather}
expressed in the global basis $\left\{\vec{e}_{1},\vec{e}_{2},\vec{n}\right\}$. The vectors $\vec e_{i,1},\vec e_{i,2},\vec e_{i,3}$ form the basis of the right-handed coordinate system for the spin at site $\vec R_i$, with the spin along the $\vec e_{i,3}$ direction in the classical ground state. To calculate the magnon spectrum, 
we apply the Holstein--Primakoff transformation \cite{Holstein_Primakoff_Trafo} in its linearised version, 
\begin{gather}
    \oS_{i,1} = \sqrt{\frac{S}{2}}\left(\oa_i + \oad_i \right)\,,  \qquad \oS_{i,2} = -i \sqrt{\frac{S}{2}}\left(\oa_i -\oad_i \right) \, ,\notag\\ \oS_{i,3} = S - \on{i}\, ,\label{eq:lin_spin_wave_def}
\end{gather}
with $\oS_{i,\alpha}=\ovS_i\cdot\vec e_{i,\alpha}$ using the coordinate system defined in Eq.~\eqref{eq:eigenvectors}, and choosing the operators corresponding to the right-handed ordering of the eigenvectors. $\left[\oa_i,\oad_j\right]=\delta_{ij}$ are bosonic creation and annihilation operators.
This so-called linear spin-wave approximation is an expansion of $\sqrt{1 - \hat a^\dagger \hat a/(2S)}$ for small magnon occupation numbers $\left<\hat a^\dagger \hat a\right>$ compared to the spin length $2S$. We want to emphasize that since the spin waves are delocalized as will be seen below, this only means that the total number of magnons must be smaller than the total spin $\sum_{\vec{k}}\left<\on{\vec k}\right>\ll 2NS$ with $N$ being the number of lattice sites. Due to this, larger magnon numbers are possible without violating the linear spin-wave condition. 

After performing Fourier transformation in real space, the spin-wave Hamiltonian takes the form
\begin{align}
    \oH_{\textrm{SW}} = \sum_{\vec k \in \text{BZ}}\left[ \omega_{\vec k} \on{\vec k} + \mu_{\vec k}\oa_{\vec k}\oa_{-\vec k} + \mu_{\vec k}^\ast \oad_{\vec k}\oad_{-\vec k}\right]\, 
    \label{eq:Ham_with_aa_dag}
\end{align}
where $\vec k$ runs over the Brillouin zone (BZ) of the system. Note that the Fourier transformation can only be used to diagonalize the Hamiltonian since the spin spiral is harmonic and can be described by a single wave vector $\vec{q}_{0}$. For anharmonic spirals, the Brillouin zone would have to be reduced based on the period of the spiral, and multiple bands would appear.

The parameters in Eq.~\eqref{eq:Ham_with_aa_dag} are given by 
\begin{align}
\begin{split}
    \mu_{\vec k} =& \frac{S}{4}s_\vartheta^2 \left[J_{\vec k} - \frac 1 2 \left(J_{\vec k - \vec q_0} + J_{\vec k + \vec q_0}\right) + \right. \\& \phantom{\frac{S^2} 2 s_\vartheta^2 [J_{\vec k}}+\left. \frac i 2 \left(D_{\vec k- \vec q_0}^{\vec{n}} - D_{\vec k + \vec q_0}^{\vec{n}}\right)\right]\, ,\label{eq:Full_Mu_k}
\end{split}
\end{align}
and
\begin{align}
     \omega_{\vec k}=&2 \mu_{\vec k} - S \left(J_{\vec q_0} + i D_{\vec q_0}^{\vec{n}}\right)+\notag\\&+\frac {S} 2 \left(1-c_{\vartheta}\right)\left(J_{\vec k + \vec q_0} + i D_{\vec k + \vec q_0}^{\vec{n}}\right)+\label{eq:DefSmallOmega}\\
    &+ \frac {S} 2 \left(1+c_{\vartheta} \right)\left(J_{\vec k - \vec q_0} - i D_{\vec k - \vec q_0}^{\vec{n}}\right) \, , \notag
    \end{align}
where we introduced $\sin \vartheta = s_\vartheta$ and $\cos \vartheta = c_\vartheta$ to shorten the expressions. 
In the determination of $\mu_{\vec k}$ and $\omega_{\vec k}$ we used the second solution in Eq.~(\ref{eq:Theta_min}) and substituted the parts containing the magnetic field through $A_{\vec q_0}$ and $\cos{(\vartheta)}$. In case of $\sin(\vartheta) = 0$ being the appropriate ground state, one would end up with $\mu_{\vec k} = 0$. This corresponds to the field-polarized ferromagnetic case and would already diagonalise the Hamiltonian in the Fock space for magnons with wave vector $\vec k$, with energies equal to
\begin{align}
     \omega_{\vec k} = S (J_{\vec k} - i D_{\vec k}^{\vec{n}}) - S J_{\vec 0} + \muB S\, .\label{eq:FMspectrum}
\end{align}

It can be shown that $\mu_{-\vec k} = \mu_{\vec k}$ and $\omega_{-\vec k} \neq \omega_{\vec k}$ by using
\begin{align}
\begin{split}
    J_{\vec k+ \vec q} &\quad \overset{\vec k \to - \vec k}{\longrightarrow}\quad J_{-\vec k + \vec q} = \phantom{-}J_{\vec k- \vec q}\, ,\\ D^{\vec{n}}_{\vec k + \vec q} &\quad \overset{\vec k \to - \vec k}{\longrightarrow}\quad D^{\vec{n}}_{-\vec k +\vec q} = -D^{\vec{n}}_{\vec k- \vec q}\, .
\end{split}
\end{align}
The final step in the diagonalization procedure is to perform a Bogoliubov transformation by introducing new bosonic operators $\oal_{\vec k}$ which are connected to the original operators $\oa_{\vec k}$ by the Bogoliubov matrix 
\begin{align}
    \begin{pmatrix}\oa_{\vec k}\\\oad_{-\vec k}\end{pmatrix} &=   \begin{pmatrix}
        u_{\vec k} &  v_{\vec k}\\ v_{\vec k}^\ast & u_{\vec k}^\ast 
    \end{pmatrix}\begin{pmatrix}\oal_{\vec k}\\\oald_{-\vec k}\end{pmatrix}\, , & |u_{\vec k}|^2 - |v_{\vec k}|^2& = 1\, .
    \label{eq:DefBogoliubovTrafo}
\end{align}
After performing the Bogoliubov transformation, the Hamiltonian takes the form 
\begin{align}
    \oH = \sum_{\vec k\in \text{BZ}}  \Omega_{\vec k} \onal{\vec k} \, ,
\end{align}
which is diagonal in the Fock space of the new magnons $\oal_{\vec k}$. The Bogoliubov parameters take the values 
\begin{align}
    u_{\vec k} &= \sqrt{\frac{\omega_{\vec k} + \Omega_{-\vec k}}{\Omega_{\vec k} + \Omega_{-\vec k}}}\, ,& v_{\vec k} &= -e^{-i \varphi_{\mu_{\vec{k}}}} \sqrt{\frac{\omega_{\vec k} - \Omega_{\vec k}}{\Omega_{\vec k} + \Omega_{-\vec k}}}\, ,\label{eq:Bogoliubov_Parameters}
\end{align}
with $\varphi_{\mu_{\vec{k}}}$ being the complex phase of $\mu_{\vec k}$. The frequency of a magnon with wave vector $\vec k$ is given by
\begin{widetext}
\begin{align}
     \Omega_{\vec k} &= \frac 1 2 \left[\left(\omega_{\vec k} - \omega_{-\vec k}\right) + \sqrt{\left(\omega_{\vec k} + \omega_{-\vec k}\right)^2 +  4|\mu_{\vec k}|^2}\right] =  S c_\vartheta \left[J_{\vec k + \vec q_0} - J_{\vec k - \vec q_0} + i \left(D^{\vec{n}}_{\vec k + \vec q_0} + D^{\vec{n}}_{\vec k - \vec q_0}\right)\right]+ \notag\\
			&+ S\sqrt{-\left(J_{\vec  q_0} + i D^{\vec{n}}_{\vec q_0}\right)+ \frac 1 2 \left(J_{\vec k + \vec q_0} + J_{\vec k - \vec q_0}\right) + \frac i 2 \left(D_{\vec k+ \vec q_0}^{\vec{n}} - D_{\vec k - \vec q_0}^{\vec{n}}\right)}\times\label{eq:Frequency}\\
			&\times \sqrt{s_\vartheta^2 J_{\vec k} +  \frac{c_\vartheta^2}{2}\left[J_{\vec k + \vec q_0} + J_{\vec k - \vec q_0} +  i  \left(D_{\vec k+ \vec q_0}^{\vec{n}} - D_{\vec k - \vec q_0}^{\vec{n}}\right)\right]-\left(J_{\vec  q_0} + i D^{\vec{n}}_{\vec q_0}\right)}\notag\, .
\end{align}
\end{widetext}
Using $\omega_{\vec k} \neq \omega_{-\vec k}$ it becomes clear that $\Omega_{\vec k} \neq \Omega_{-\vec k}$ when the spin spiral is not planar, i.e., for $\cos\vartheta\neq 0$. This is known as non-reciprocal magnon propagation. While the non-reciprocal propagation is often connected to the presence of the DMI, which is indeed required for it in the ferromagnetic limit of Eq.~\eqref{eq:FMspectrum}, in this case the symmetry between $\vec{k}$ and $\vec{-k}$ magnons is broken by the rotational sense of the spin spiral characterized by $\vec{q}_{0}$, and the presence of a finite net magnetic moment that breaks time-reversal symmetry.

The spin-wave dispersion in Eq.~\eqref{eq:Frequency} is equivalent to the zero-temperature limit of the dispersion relation derived in Ref.~\cite{SpinSpiralDispersion_Michael} for $S=1/2$. 

\subsection{Squeezing parameters}
\label{subsec:2c}
\begin{figure}
	\begin{center}
			\includegraphics[width = \columnwidth]{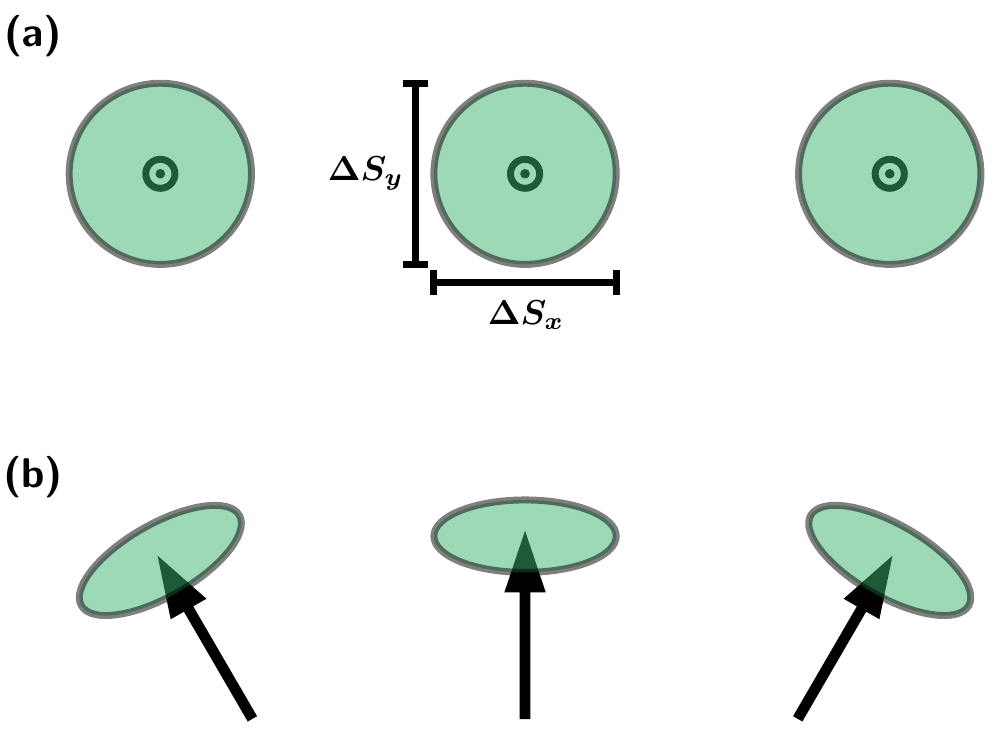}
	\end{center}
	\caption{Variances of the spin components orthogonal to the ground state direction. (a) Ferromagnetic ordering with no squeezing in the variances of the $x$ and $y$ component of the spins. Variances at different sites lie in the same plane. (b) Spin spiral ordering with squeezed variances in orthogonal directions. Variances lie in different planes depending on the site and the corresponding local frame Eq.~\eqref{eq:eigenvectors}.}
	\label{fig:2}
\end{figure}
The Bogoliubov transformation in Eq.~\eqref{eq:DefBogoliubovTrafo} connects the magnon operators with wave vectors $\vec{k}$ and $-\vec{k}$. This is analogous to the situation in ferromagnets~\cite{Non_Integer_Spin_Akash} and in antiferromagnets~\cite{AFM_Squeezing_Akash}, which are included in the present formalism for spin-spiral wave vectors $\vec{q}_{0}=\vec{0}$ and $\vec{q}_{0}=\vec{b}/2$, where $\vec{b}$ is a basis vector of the atomic reciprocal lattice, respectively. Squeezing between $\vec{k}$ and $-\vec{k}$ is described by the two-mode-squeezing operator 
\begin{align}
    \hat S_2(r_{\vec k}) = \exp\left(r_{\vec k}^\ast \oa_{\vec k}\oa_{-\vec k} -r_{\vec k} \oad_{\vec k} \oad_{-\vec k}\right)\, .\label{eq:squeezingoperator}
\end{align}
The degree of squeezing in a system can be given in terms of the absolute value of the complex squeezing parameter $r_{\vec k}$. In a biaxial ferromagnet, squeezing can be imagined as a reduction of the standard deviation of the spin component along the hard axis and a simultaneous increase of the standard deviation of the spin component along the intermediate axis. The product of the standard deviations, as limited by the Heisenberg uncertainty principle, remains conserved. In a conical spin spiral the squeezing also describes the different values of the standard deviations of the spin components perpendicular to the equilibrium direction, but the situation is more complicated since these directions are defined in the local coordinate system of Eq.~\eqref{eq:eigenvectors} which changes from site to site, as can be seen in Fig.~\ref{fig:2}. The phase $\varphi_{r_{\vec{k}}}$  of $r_{\vec k}$ determines along which direction the standard deviation is reduced or enhanced. However, the value of the phase is not gauge-invariant, i.e., it depends on the choice of $\vec e_{i,1}$ and $\vec e_{i,2}$ in Eq.~\eqref{eq:eigenvectors} which may be freely chosen as long as the right-handed orientation of the frame is conserved. 

The product of the squeezing operators over the $\vec k$ vectors in half of the Brillouin zone $\prod_{\vec{k}}\hat S_{2}(r_{\vec k})$ transforms the classical vacuum state which is destroyed by each original magnon operator $\oa_{\vec k}$ to the approximate quantum or squeezed vacuum of linear spin-wave theory which is destroyed by the magnon eigenstates $\oal_{\vec k}$. Note that the squeezing operator creates pairs of the original magnons with opposite wave vectors, similarly to Cooper pairs in Bardeen--Cooper--Schrieffer theory; the difference is that higher magnon occupations are also present in the squeezed vacuum due to the bosonic commutation relations. Due to the construction of the bosonic Fock space, the original magnon states are highly entangled in the squeezed vacuum~\cite{AFM_Squeezing_Akash}.

The connection between the vacua implies that the transformation of the operators has to follow
\begin{align}
    \begin{split}
        \oal_{\vec k} &=\hat S_2(r_{\vec k}) \oa_{\vec k}\hat S^{-1}_2(r_{\vec k}) = \\
        &= \cosh(|r_{\vec k}|) \oa_{\vec k} + e^{i \varphi_{r_{\vec{k}}}}  \sinh(|r_{\vec k}|) \oad_{-\vec k}\, .
    \end{split}
\end{align}
The connection between the squeezing parameter and the matrix elements of the Bogoliubov matrix in Eq.~\eqref{eq:DefBogoliubovTrafo} is given by
\begin{align}
    u_{\vec k} &=  \cosh(|r_{\vec k}|) \, , & v_{\vec k}&= -e^{i \varphi_{r_{\vec{k}}}}  \sinh(|r_{\vec k}|) \, .
\end{align}
Using this equation and Eq.~\eqref{eq:Bogoliubov_Parameters}, one can determine the squeezing parameter for a mode with wave vector $\vec k$ in a conical spin spiral,
\begin{align}
    \tanh(|r_{\vec k}|) &= \sqrt{\frac{\omega_{\vec k} - \Omega_{\vec k}}{\omega_{\vec k} + \Omega_{-\vec k}}}\, , & \varphi_{r_{\vec{k}}} &= - \varphi_{\mu_{\vec{k}}}\, ,\label{eq:Squeez_Param}
\end{align}
constituting the central result of this work. 

Using Eq.~\eqref{eq:Frequency} one can show that
\begin{align}
\omega_{\vec k} \mp \Omega_{\pm\vec k} &= \omega_{-\vec k} \mp \Omega_{\mp\vec k} \, . \label{eq:Omega_Condition}
\end{align}
Together with Eq.~\eqref{eq:Squeez_Param} this leads to 
\begin{align}
    \tanh(|r_{-\vec k}|) &= \sqrt{\frac{\omega_{-\vec k} - \Omega_{-\vec k}}{\omega_{-\vec k} + \Omega_{\vec k}}}  &\overset{\text{Eq.\eqref{eq:Omega_Condition}}}{=} \sqrt{\frac{\omega_{\vec k} - \Omega_{\vec k}}{\omega_{\vec k} + \Omega_{-\vec k}}}   = \notag \\&= \tanh(|r_{\vec k}|)\,, 
\end{align}
meaning that the squeezing parameter is invariant under wave-vector inversion despite the non-reciprocal magnon propagation. In the present choice of gauge, the phase $\varphi_{r_{\vec{k}}} = - \varphi_{\mu_{\vec{k}}}$ is also invariant since $\mu_{\vec k}$ is even under inverting the wave vector. This is explained by the fact that the two-mode-squeezing operator in Eq.~\eqref{eq:squeezingoperator} assigns a single squeezing parameter to the pair of magnon operators at wave vectors $\vec k$ and $-\vec k$. It is also worth noting that the squeezing vanishes for $\mu_{\vec{k}}=0$, since this leads to $\Omega_{\vec{k}}=\omega_{\vec{k}}$, for example in the collinear polarized state with the spectrum given in Eq.~\eqref{eq:FMspectrum}. 

It is also important to note that the squeezing parameter diverges as $\vec{k}\to \vec{0}$, and is not defined for the uniform mode $\vec k = \vec 0$. For this mode $\omega_{\boldsymbol{0}} = 2 \mu_{\vec{0}}$ and $\Omega_{\boldsymbol{0}} = 0$ which makes it impossible to diagonalize the Hamiltonian using a Bogoliubov transformation fulfilling Eq.~\eqref{eq:DefBogoliubovTrafo}. In the formalism of non-Hermitian eigenvalue equations, this is known as an exceptional point; note that while the spin-wave Hamiltonian is Hermitian, the equation of motion is enforced to be non-Hermitian by the bosonic commutation relations~\cite{Flynn2020}. 

The divergence of the squeezing parameter at $\vec k = \vec 0$ was also found in isotropic antiferromagnets in Ref.~\cite{AFM_Squeezing_Akash}, and is connected to the Goldstone mode of the system. In a commensurate spin structure such as a collinear antiferromagnet, the Goldstone mode related to the global spin rotation may be lifted by an arbitrarily weak anisotropy term. However, for the incommensurate spin spirals discussed here, the Goldstone mode is related to the translation of the spiral along the wave vector direction $\vec{q}_{0}$, and including an anisotropy term in the plane of the spiral would only distort its shape but would not obstruct its translation, unless the anisotropy term is strong enough to lock the spiral into a commensurate state. This implies that the magnon modes with low wave vectors in conical spin spirals are always very strongly squeezed.

\begin{figure}
	\begin{center}
			\includegraphics[width = \columnwidth]{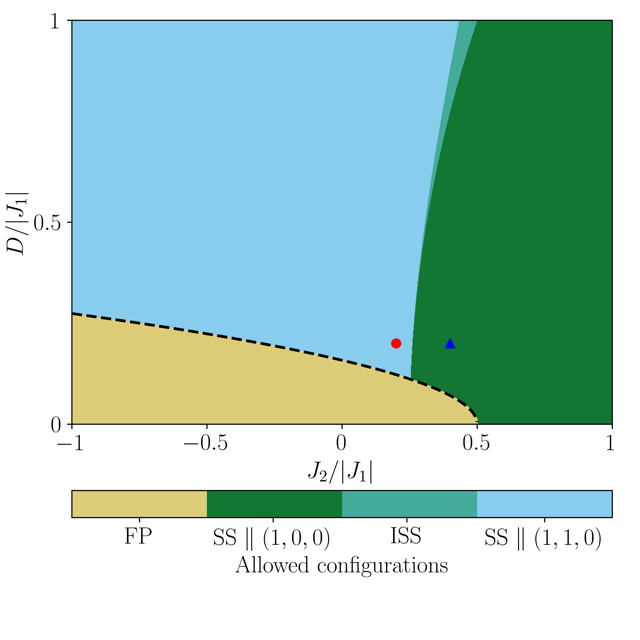}
	\end{center}
	\caption{Phase diagram of the classical ground state of a two-dimensional square lattice magnet as a function of NNN Heisenberg coupling $J_{2}$ and DMI $D$, displaying field-polarized (FP), conical spin spiral (SS) and intermediate spin spiral (ISS) configurations with different orientations of the wave vector $\boldsymbol{q}_{0}$. We fixed the field energy to be $\muB = 0.025|J_1|$. The red dot and the blue triangle correspond to $J_2 = 0.2 |J_1|$, $D = 0.2  |J_1|$, and $J_{2}=0.4 |J_1|$, $D = 0.2  |J_1|$, respectively. The black dashed line is the analytic boundary of the FP state given by Eq.~\eqref{eq:ParabolicShape}.}
	\label{fig:3}
\end{figure}
\begin{figure}
	\begin{center}
			\includegraphics[width = \columnwidth]{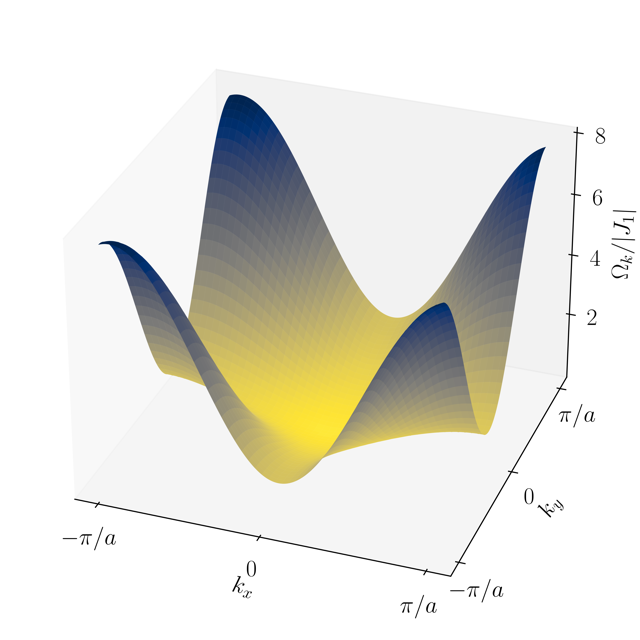}
	\end{center}
	\caption{Energy dispersion, in units of $|J_1|$, for magnons with wave vector $\vec k$ 
 when the system is in the classical SS configuration along $(1,0,0)$ with $\muB = 0.025|J_1|$, $J_2 = 0.4 |J_1|$ and $D = 0.2  |J_1|$.}
	\label{fig:4}
\end{figure}

Since the squeezing parameter is related to, although not simply proportional to, the parameter $\mu_{\vec{k}}$ in Eq.~\eqref{eq:Full_Mu_k}, analyzing this equation will give a qualitative understanding of the squeezing parameter. The value of $\mu_{\vec{k}}$ decreases as $\sin\vartheta$ is decreased, i.e., as the external magnetic field closes the cone angle and drives the system into the collinear state. For $\vec{q}_{0}=0$ describing a ferromagnetic alignment, the squeezing vanishes as no anisotropy is present, as already discussed in Ref.~\cite{Non_Integer_Spin_Akash}. With increasing values of $\vec{q}_{0}$, $\mu_{\vec{k}}$ increases as the Heisenberg interactions and the DMI start contributing to the squeezing. This is relevant in particular for magnon wave vectors $\lvert\vec{k}\rvert\ll \lvert\vec{q}_{0}\rvert$, and the squeezing is expected to decrease for larger $\vec{k}$. However, apart from this qualitative decrease, the condition $\mu_{\vec{k}}=0$ implying a vanishing squeezing parameter defines a subspace of codimension one in reciprocal space, i.e., a curve in two dimensions and a surface in three dimensions. Magnons with opposite wave vectors located on this subspace are not present in the squeezed vacuum, and thus are exempt from the entanglement.

\section{Application to an example system}
\label{sec:3}
\begin{figure*}
    \subfloat[\label{fig:5a}]{%
          \includegraphics[width=\columnwidth]{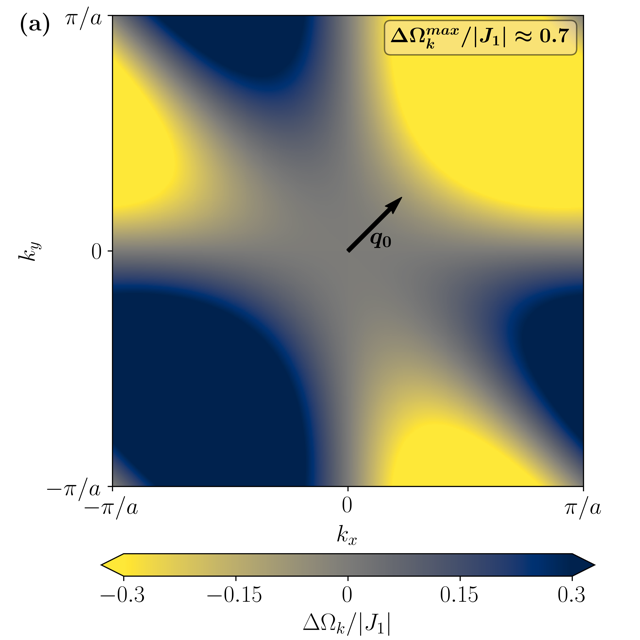}%
          
        }\hspace*{\fill}%
        \subfloat[\label{fig:5b}]{%
          \includegraphics[width=\columnwidth]{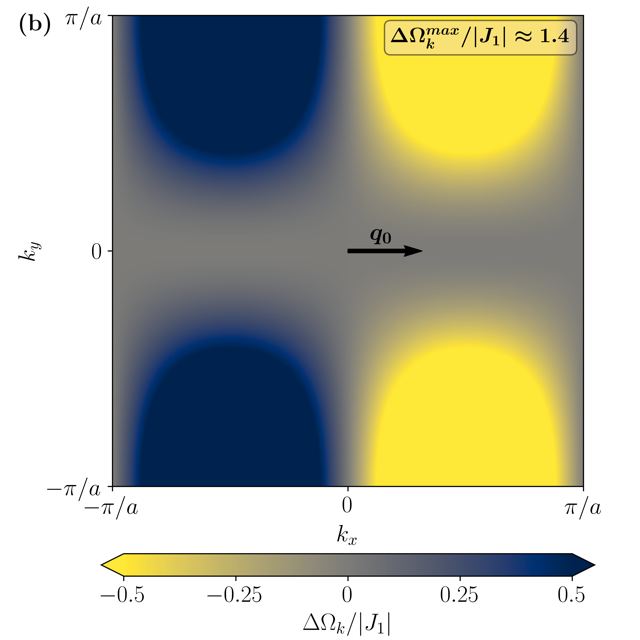}%
        }
	\caption{
	Magnon non-reciprocity $\Delta \Omega_{\vec k} = \Omega_{\vec k} - \Omega_{-\vec k}$ for different directions of the spin spiral wave vector $\vec{q}_{0}$. 
    (a) SS along $(1,1,0)$ with $J_2 = 0.2 |J_1|$ and $D = 0.2 |J_1|$. The colorbar is capped at $|\Delta \Omega_{\vec k}| = 0.3 |J_1|$ with maximal energy difference $\Delta \Omega_{\vec k}^{\text{max}} = 0.7 |J_1|$.
    (b) SS along $(1,0,0)$ with $J_2 = 0.4  |J_1|$ and $D = 0.2 |J_1|$. The colorbar is capped at $ |\Delta \Omega_{\vec k}| = 0.5 |J_1|$ with maximal energy difference $ \Delta \Omega_{\vec k}^{\text{max}} = 1.4 |J_1|$. 
    The arrows show the direction of 
    $\vec q_0$ for each spin spiral but not its proper length.
    Both figures use $\muB = 0.025  |J_1|$ and $S = 1$.}
	\label{fig:5}
\end{figure*}
As an example system, we investigate a 2D square lattice lying in the $xy$ plane. We consider a NN ferromagnetic Heisenberg interaction $J_{1}<0$, and express all other parameters in units of $\lvert J_{1}\rvert$. Besides the field energy $\muB$, we take into account NNN Heisenberg interaction $J_{2}$ and NN DMI of strength $D$. We assume the square lattice to be on a non-magnetic substrate with $C_{4\textrm{v}}$ symmetry. The substrate is necessary for breaking the inversion symmetry required for a finite DMI. The interactions are sketched in Fig.~\ref{fig:B1} in Appendix~\ref{sec:appendixb}. 

Generally, both an antiferromagnetic NNN Heisenberg term $J_{2}>0$ and the DMI $D$ could stabilize a spin spiral when competing with $J_{1}$. As will be discussed below, in the present system only ferromagnetic and row-wise antiferromagnetic states are stabilized in the absence of DMI. The $J_{2}$ term enables to tune the wave vector of the spin spiral in the whole range between the ferromagnetic and the antiferromagnetic limits, while the DMI would only be able to open a maximum of $\pi/2$ angle between neighbouring spins. Another motivation for taking into account $J_{2}$ is that for only NN Heisenberg and DMI terms, the magnon dispersion would be reciprocal in the conical spin spiral state, $\omega_{\vec{k}}=\omega_{-\vec{k}}$ in Eq.~\eqref{eq:DefSmallOmega}. This situation has to be avoided since it would be related to the choice of parameters, not to the symmetry of the system. 

The Fourier transform of the interactions are given by 
\begin{align}
    J_{\vec k} = 2J_1\left[\cos(a k_x) + \cos(a k_y)\right] + 4 J_2\cos(a k_x)\cos(a k_y)\, ,\label{eq:Jq}
\end{align}
and 
\begin{align}
     \vec D_{\vec k} &= - 2 i D \begin{pmatrix}
 		- \sin(a k_y) \\ \sin(a k_x)\\ 0
 	\end{pmatrix}\, ,\label{eq:Dq}
\end{align}
where $a$ is the lattice constant.
\begin{figure*}
        \subfloat[\label{fig:6a}]{%
          \includegraphics[width=\columnwidth]{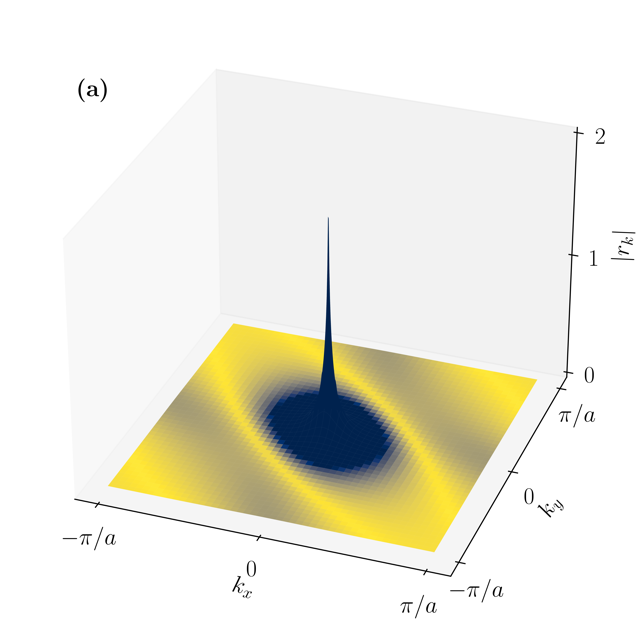}%
        }\hspace*{\fill}%
        \subfloat[\label{fig:6b}]{%
          \includegraphics[width=\columnwidth]{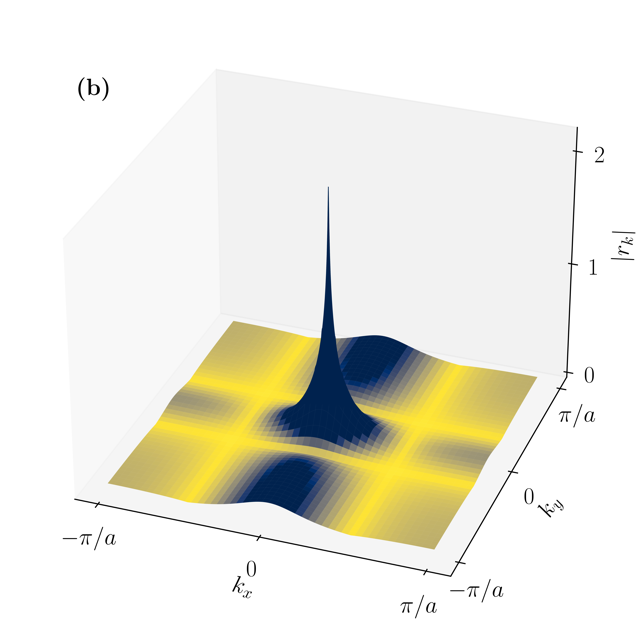}%
        }
	\caption{Absolute value of the squeezing parameter $r_{\vec k}$ for different classical ground states.
    (a) SS along $(1,1,0)$ with $J_2 = 0.2  |J_1|$ and $D = 0.2 |J_1|$. The color scale is capped at $|r_{\vec k}| = 0.01$, such that finer details get revealed.
    (b) SS along $(1,0,0)$ with $J_2 = 0.4  |J_1|$ and $D = 0.2 |J_1|$. 
    The color scale is capped at $|r_{\vec k}| = 0.1$.
    In both cases the squeezing diverges when approaching the uniform $\vec k = 0$ mode. 
    Both figures use $\muB = 0.025  |J_1|$.}
	\label{fig:6}
\end{figure*} 

As discussed in Sec.~\ref{subsec:2b}, for each spiral wave vector $\vec{q}_{0}$ we select the cone axis $\vec{n}$ to be antiparallel to $\vec D_{\vec q_0}$, leading to 
\begin{align}
    D_{\vec q_0}^{\vec{n}} &= +2i|D| \sqrt{\sin^2(a q_0^y) +\sin^2(a q_0^x)}\, .\label{eq:Dqabs}
\end{align}
Note that the $\vec{n}$ direction always lies in the $xy$ plane. 

With Eqs.~\eqref{eq:Jq} and \eqref{eq:Dqabs} we can now determine the minima of the classical energy using Eq.~\eqref{eq:q0_min}. An analytical calculation of the extrema detailed in Appendix~\ref{sec:appendixb} yields four kinds of possible configurations, which are displayed in the phase diagram in Fig.~\ref{fig:3}. 

The first one is the field-polarized (FP) configuration, corresponding to a collinear alignment of the spins along the magnetic field direction with $\vartheta \in \{0, \pi\}$. This is the ground state if the magnetic field becomes stronger than the interactions between the spins, represented by $A_{\vec q_{0}}$ in Eq.~\eqref{eq:Theta_min}, thereby closing the cone angle. The boundary of the FP configuration in the phase diagram can be estimated for small field strengths by 
\begin{align}
    \lvert D\rvert \leq \sqrt{-\muB(J_1 + 2 J_2)}\, 
    \label{eq:ParabolicShape}
\end{align}
for $J_1 + 2 J_2<0$. This is denoted by the black dashed line in Fig.~\ref{fig:3}. As expected, the field-polarized region becomes more extended as the magnetic field is increased. 

The other three phases correspond to conical spin spiral (SS) configurations differing in the direction of the wave vector $\vec{q}_{0}$, on which the NNN Heisenberg coupling $J_{2}$ has the strongest effect. We get $\vec q_0 \parallel \vec e_{x/y}$ for large values of $J_{2}/\lvert J_{1}\rvert$. This corresponds to ferromagnetic rows along the $\vec e_{y/x}$ direction while along the perpendicular direction the spins are rotating. For small $J_{2}/\lvert J_{1}\rvert$ the wave vector lies along the diagonal of the square lattice with $q_0^x = \pm q_0^y$. The two regions are connected by the intermediate spin spiral (ISS) configuration where $\vec{q}_{0}$ continuously rotates from the $(1,0,0)$ direction along the axis towards the $(1,1,0)$
along the diagonal, given by the expression
\begin{align*}
    \sqrt{\sin^2(aq_0^x) + \sin^2(aq_0^y)} = \frac{|D|}{2 J_2}\, .
\end{align*}
This region becomes wider for higher $D$ values. Note that in all spin spiral configurations, energetically equivalent states are found if $\vec{q}_{0}$ and the cone axis direction are transformed by the symmetries of the square lattice. 

We also discuss the configurations along the line $D=0$. For $\muB=0$, a ferromagnetic state ($\vec{q}_{0}=\vec{0}$) is formed for $J_{2}<\lvert J_{1}\rvert/2$, and the classical ground state is row-wise antiferromagnetic ($\vec{q}_{0}=\left(\pi/a,0,0\right)$) for $J_{2}>\lvert J_{1}\rvert/2$. At the point $J_{2}=\lvert J_{1}\rvert/2$, all spin spirals with wave vectors along the $\left(1,0,0\right)$ direction are degenerate. This demonstrates that the DMI is necessary for finding a unique spin spiral ground state. Under applying a magnetic field, the ferromagnetic and the antiferromagnetic configurations transform into a tilted collinear and a spin-flop state, respectively. The magnetic field also selects the $\vec{q}_{0}=\vec{0}$ wave vector as the energetically preferred one at $J_{2}=\lvert J_{1}\rvert/2$. These configurations are difficult to denote in the two-dimensional phase diagram since they are restricted to a line. However, they can be included in the SS configuration with $\vec{q}_{0}\parallel\left(1,0,0\right)$, for specific values of the wave vector.

We investigate the magnon spectrum and the squeezing in two points of the phase diagram, $J_2 = 0.2 |J_1|$, $D = 0.2 |J_1|$ with $\vartheta \approx 0.4 \pi$ and $a  |\vec q_0| \approx 0.1\pi$ along $(1,1,0)$, and $J_2 = 0.4 |J_1|$, $D = 0.2 |J_1|$ with $\vartheta \approx 0.45 \pi$ and $a |\vec q_0| = \pi/4$ along $(1,0,0)$, denoted by a red dot and a blue triangle in Fig.~\ref{fig:3}, respectively.

The magnon dispersion for the blue triangle in Fig.~\ref{fig:3} is shown in Fig.~\ref{fig:4}, while the magnon non-reciprocity given by
\begin{align}
     \Delta \Omega_{\vec k} &= \left(\Omega_{\boldsymbol k} - \Omega_{-\boldsymbol k}\right) = \notag \\
    &=2 S c_{\vartheta} \left[J_{\vec k + \vec q_0} - J_{\vec k - \vec q_0} + i \left(D_{\vec k + \vec q_0}^{\vec{n}} + D_{\vec k - \vec q_0}^{\vec{n}}\right)\right]\, 
    \label{eq:EnergyDiff}
\end{align}
is illustrated in Fig.~\ref{fig:5}(a) and (b) for the two states mentioned above.

The spectrum in Fig.~\ref{fig:4} displays a Goldstone mode with zero frequency $\vec{k}=\vec{0}$, which arises because translating the spin spiral along $\vec{q}_{0}$ costs no energy, as discussed in Sec.~\ref{subsec:2c}. Overall, the magnon frequencies increase when moving away from the center of the Brillouin zone, but there is an asymmetry between wave vectors $\vec{k}$ and $-\vec{k}$ which is easier to see in Fig.~\ref{fig:5}. This asymmetry reaches high values of the order of $\lvert J_{1}\rvert$ away from the center of the Brillouin zone. Comparing Figs.~\ref{fig:5}(a) and (b), it can be concluded that the energy difference completely vanishes orthogonal to $\vec q_0$ as well as along the lattice vectors $\vec e_{x/y}$. The disappearance of the non-reciprocity perpendicular to $\vec q_0$ is enforced by the $C_{4\textrm{v}}$ symmetry of the system when the spin spiral wave vector lies in a mirror plane. The vanishing of $\Delta\Omega_{\vec{k}}$ along the main axes appears to be specific to the choice of interaction parameters, and may be lifted if interactions with further neighbors are taken into account. Even for a generic direction of $\vec{q}_{0}$ (e.g., in the ISS) and for a spin model containing more interaction terms, there must be a line crossing the whole Brillouin zone along which the non-reciprocity vanishes, since $\Delta\Omega_{\vec{k}}$ is continuous and by definition has to change sign when reversing the wave vector direction. Overall, magnons travelling along $\vec{q}_{0}$ opposed to along $-\vec{q}_{0}$ seem to have a lower energy, therefore they are easier to excite.

The squeezing parameters from Eq.~\eqref{eq:Squeez_Param} are displayed in Fig.~\ref{fig:6} for the two considered spin spiral states. This displays the patterns discussed in Sec.~\ref{subsec:2c}. The squeezing, in contrast to the dispersion, is always symmetric between $\vec{k}$ and $-\vec{k}$. In particular, the squeezing parameter respects a $C_{2\textrm{v}}$ symmetry, as expected by reducing the $C_{4\textrm{v}}$ symmetry by the direction of the spin spiral wave vector which is located in a mirror plane. The squeezing parameter diverges as the Goldstone mode $\vec{k}=\vec{0}$ is approached. It generally decreases towards the boundary of the Brillouin zone, similarly to what was observed for the antiferromagnetic configuration in Ref.~\cite{AFM_Squeezing_Wuhrer}. However, at the brightest yellow curves inside the atomic Brillouin zone the parameter $\mu_{\vec k}$ in Eq.~\eqref{eq:Full_Mu_k}, and consequently the squeezing $r_{\vec{k}}$, vanish. For the wave vector along the $\left(1,1,0\right)$ direction in Fig.~\ref{fig:6}(a), these curves are almost perpendicular to $\vec{q}_{0}$. For $\vec{q}_{0}\parallel\left(1,0,0\right)$ in Fig.~\ref{fig:6}(b), the curves with vanishing squeezing are almost parallel to the $x$ and $y$ directions, which are parallel and perpendicular to $\vec{q}_{0}$, respectively. Here, the squeezing does vanish for certain wave vectors perpendicular to $\vec{q}_{0}$. The distance of the curves with vanishing $r_{\vec{k}}$ from the center
of the Brillouin zone depends on the magnitude of $\vec{q}_{0}$, but the quantitative relationship between these quantities appears to be dependent on the model parameters. 

\section{Conclusion}

We calculated the squeezing of magnons in a conical spin spiral state. Within linear spin-wave theory, the ground state of the system may be described by a vacuum where pairs of magnons with wave vectors $\vec{k}$ and $-\vec{k}$ undergo squeezing. Although the spin spiral structure together with a finite net magnetization leads to a non-reciprocal propagation of magnons, the squeezing parameter is symmetric under reversing the direction of the wave vector since it describes a pair of magnons with opposite wave vectors.

The degree of squeezing in the spin spiral state interpolates between the ferromagnetic limit, where it vanishes due to the absence of an anisotropy term, to the antiferromagnetic limit, where it is exchange-dominated, by changing the wave vector $\vec{q}_{0}$ of the spiral or by closing the cone angle. The squeezing is found to diverge when approaching the $\vec{k}=\vec{0}$ Goldstone mode, and this divergence is not possible to be removed by magnetic anisotropy in incommensurate spin spiral states, in contrast to the commensurate ferromagnetic or antiferromagnetic states. The squeezing parameter is qualitatively found to decrease away from the center of the Brillouin zone, but it exactly vanishes for wave vectors located on certain curves in two-dimensional and on surfaces in three-dimensional systems.

The results were illustrated on a two-dimensional square lattice taking nearest-neighbor and next-nearest-neighbor Heisenberg and Dzyaloshinsky--Moriya interactions into account. We identified the regime in the parameter space where the conical spin spiral state is stable, and discussed how the preferred orientation of the spin spiral wave vector is changing while varying the interactions. The direction of the wave vector is found to be reflected in the non-reciprocity of the magnon dispersion and the squeezing parameter in the Brillouin zone.

The squeezing of magnons describes the decrease in the standard deviation of one spin component at the cost of increasing the standard deviation in the conjugate spin component in the plane perpendicular to the magnetization direction in the classical ground state. As described in Ref.~\cite{Udvardi2003}, this effect is intrinsically related to the classical concept of elliptic spin-wave polarization, where the spins precess on an elliptic path around their equilibrium direction. The analogy relies on the equivalence of the calculation of magnon frequencies and eigenvectors in quantum and classical linear spin-wave theory. This makes it possible to assess certain signatures of squeezing in classical observables~\cite{AFM_Squeezing_Akash}: for example, the elliptic polarization of spin waves is reflected in their linewidth in resonance experiments~\cite{Rozsa2018}, while the different components of transversal spin correlations are accessible in spin-polarized electron and neutron scattering experiments. Importantly, the squeezing also leads to a decrease in the longitudinal spin component in the quantum limit, which is not observed in the classical case. This enables its detection through longitudinal spin oscillations, for example via the measurement of the light polarization rotation after the optical excitation of magnon pairs with opposite wave vectors in Ref.~\cite{Bossini2019}. The creation of such magnon pairs in the conical spin spiral states discussed here would be particularly intriguing, since the magnons travelling along opposite directions possess different frequencies due to the non-reciprocity, which might be used to spatially separate them to take advantage of the entanglement encoded in their common squeezing parameter. 

\section*{Acknowledgment} This work was financially supported by the Deutsche Forschungsgemeinschaft (DFG, German Research Foundation) via the Collaborative Research Center SFB 1432 (project no.~425217212), by the National Research, Development, and Innovation Office (NRDI) of Hungary under Project Nos. K131938 and FK142601, by the Ministry of Culture and Innovation and the National Research, Development and Innovation Office within the Quantum Information National Laboratory of Hungary (Grant No. 2022-2.1.1-NL-2022-00004), and by the Young Scholar Fund at the University of Konstanz. 

\appendix
    \section{Calculation of the magnon 
    dispersion\label{sec:appendixa}}
    Here, we give a short derivation of the energy dispersion given in Eq.~\eqref{eq:Frequency}. We start from the Hamiltonian in Eq.~\eqref{eq:Def_Ham}, assume that the classical ground state was already determined, and at each site we apply the local eigensystem $(\vec e_{i,1}, \vec e_{i,2}, \vec e_{i,3})$ given in Eq.~\eqref{eq:eigenvectors}. We define the scalar product of the local eigenvectors with the global eigenvectors of the cone system $(\vec{e}_1, \vec{e}_2, \vec n)$ as $e_{i,m}^{\alpha} = \vec e_{\alpha}\cdot \vec e_{i,m}$ with $\alpha \in \{1,2,\vec{n}\}$, $m\in\{1,2,3\}$ and $\vec e_{\vec{n}} = \vec n$.
    
    Using this, we get the following equations for the different combinations of spin products appearing in Eqs.~(\ref{eq:Heisenberg})-(\ref{eq:Anisotropy_Zeeman}):
    \begin{align}
  		\ovS_i\cdot \ovS_j &= \sum_{m,l\in\{1,2,3\}} \left(\vec e_{i,m}\cdot \vec e_{j,l} \right)\oS_{i,m}\oS_{j,l} \\
		\oS_i^\alpha  \oS_j^\beta &= \sum_{m,l\in\{1,2,3\}} e_{i,m}^{\alpha}  e_{j,l}^\beta\oS_{i,m}\oS_{j,l} \\
		\oS_i^\alpha &= \sum_{m\in\{1,2,3\}}e_{i,m}^{\alpha}\oS_{i,m} \, .
  	\end{align}
    This enables rewriting the Hamiltonian in the following manner:
    \begin{align}
        \oH &= \sum_{\vec R_i,\vec R_j} \ovS_i^\top \Xi_{ij}\ovS_j  +\sum_{\vec R_i}\vec \xi_i\ovS_i \,,
    \end{align}
    with 
    \begin{align}
        \begin{split}
            \Xi_{ij}^{m l} = &\frac 1 2 J_{ij} \left(\vec e_{i,m} \cdot \vec e_{j,l}\right) + \\
            +&\frac 1 2 D^{\vec{n}}_{ij}\left( e_{i,m}^{1} e_{j,l}^{2} -e_{i,m}^{2} e_{j,l}^{1} \right)\, ,
        \end{split}\label{eq:Xi_1} & i&\neq j\\
  		\Xi_{ii}^{m l} =&\,  0\, ,\label{eq:Xi_2}\\
  		\xi_i^{m\phantom{l}} =& -\muB  e_{i,m}^{\vec{n}}\, .\label{eq:Xi_3}
    \end{align}
    We perform the Holstein--Primakoff transformation given by Eq.~\eqref{eq:lin_spin_wave_def}, resulting in the linearized spin-wave Hamiltonian
    \begin{align}
        \oH &= \sum_{\vec R_i,\vec R_j} \hat{\vec a}_i^\dagger \chi_{ij}\hat{\vec a}_j \, ,
    \end{align}
     with $\hat{\vec a}_i = (\oa_i,\oa_i^\dagger)^\top$ and where we dropped constant terms.

    The components of the different $\chi_{ij}$ for $i\neq j$ are connected to the $\Xi_{ij}$ and $\xi_{i}$ parameters as follows:
    \begin{align}
        \chi_{ij}^{a^\dagger a}  &= \frac{S }{2} \left[\Xi_{ij}^{11} + \Xi_{ij}^{22} - i \left(\Xi_{ij}^{12} - \Xi_{ij}^{21}\right)\right] =\label{eq:Chi_1}\\&=  \left(\chi_{ij}^{a a^\dagger}\right)^\ast\, ,\notag \\
		\chi_{ij}^{a a}  &= \frac{S}{2} \left[\Xi_{ij}^{11} - \Xi_{ij}^{22} - i \left(\Xi_{ij}^{12} + \Xi_{ij}^{21}\right)\right] = \\ &=\left(\chi_{ij}^{a^\dagger a^\dagger} \right)^\ast\, , \notag
    \end{align}
    and for $i = j$ one obtains
    \begin{align}
		\chi_{ii}^{a^\dagger a}  &= \frac{S}{2} \left[\Xi_{ii}^{11} + \Xi_{ii}^{22} - i \left(\Xi_{ii}^{12} - \Xi_{ii}^{21}\right)\right] -\notag \\&\phantom{=}- S \sum_j \left(\Xi_{ij}^{33} + \Xi_{ji}^{33}\right) - 2 S \Xi_{ii}^{33} -  \xi_{i}^3= \notag\\
        &= - S \sum_j \left(\Xi_{ij}^{33} + \Xi_{ji}^{33}\right) -  \xi_{i}^3\, , \\
		\chi_{ii}^{a a^\dagger}  &= \frac{S}{2} \left[\Xi_{ii}^{11} + \Xi_{ii}^{22} + i \left(\Xi_{ii}^{12} - \Xi_{ii}^{21}\right)\right] = 0\, , \\
		\chi_{ii}^{a a}  &= \frac{S}{2} \left[\Xi_{ii}^{11} - \Xi_{ii}^{22} - i \left(\Xi_{ii}^{12} + \Xi_{ii}^{21}\right)\right] =0 
    \end{align}
     Since the system is expanded around the classical ground state, the terms linear in the creation and annihilation operators must vanish.

     In the next step, we perform Fourier transformation on the creation and annihilation operators, 
     \begin{align}
         \oa_{\vec k} &= \frac 1 {\sqrt N} \sum_i e^{+ i \vec k \cdot \vec R_i} \oa_i\, ,& \oa_{i} &= \frac 1 {\sqrt N} \sum_{\vec k} e^{- i \vec k \cdot \vec R_i} \oa_{\vec k}\, ,
     \end{align}
     which leads to the Hamiltonian in Eq.~\eqref{eq:Ham_with_aa_dag}. The coefficients are then given by 
     \begin{align}
        \mu_{\vec k} &=\chi_{ii}^{aa} + \sum_{\vec \delta} \chi_{i, i+\delta}^{aa} e^{-i\vec \delta \cdot \vec k}\, ,\\
         \omega_{\vec k} &=\chi_{ii}^{a^\dagger a} + \chi_{ii}^{a a^\dagger}  + \sum_{\vec \delta} \left[\chi_{i, i + \delta}^{a^\dagger a}e^{i\vec \delta \cdot \vec k} + \chi_{i, i+\delta}^{aa^\dagger}e^{-i\vec \delta \cdot \vec k}\right] \, ,
     \end{align}
     where $\vec \delta$ is a lattice vector. Using trigonometric identities to express the products of the components of the eigenvectors $\vec{e}_{i,m}$, Fourier transforms of the interaction coefficients with wave vectors shifted by $\vec{q}_{0}$ appear in the Hamiltonian,
     \begin{align}
        J_{\vec k \mp \vec q_0} &=  \sum_{\vec \delta}J_{\vec \delta} e^{\pm i \vec \delta \cdot \vec q_0}e^{-i \vec \delta \cdot \vec k}\, \\  D^{\vec{n}}_{\vec k \mp \vec q_0} &= \sum_{\vec \delta}D^{\vec{n}}_{\vec \delta} e^{\pm i \vec \delta \cdot \vec q_0}e^{-i \vec \delta \cdot \vec k} \, .
     \end{align}
     Collecting the terms yields 
     \begin{align}
        \omega_{\vec k} = S &\left[ - c_\vartheta^2 J_0 - \frac{s_\vartheta^2}2\left(2 J_{\vec q_0} - J_{\vec k}\right) -i s_\vartheta^2 D_{\vec q_0}^{\vec{n}}  + \right. \notag\\ &
		+\frac 1 4 \left(c_{\vartheta} - 1\right)^2\left(J_{\vec k - \vec q_0} - i D_{\vec k - \vec q_0}^{\vec{n}}\right)+ \notag \\ &\left.+\frac 1 4 \left(c_{\vartheta} + 1\right)^2\left(J_{\vec k + \vec q_0} + i D_{\vec k + \vec q_0}^{\vec{n}}\right)\right] +\notag\\
		&\phantom{[}+\muB Sc_\vartheta \, ,\label {eq:Full_small_Omega_k}
     \end{align}
     and Eq.~\eqref{eq:Full_Mu_k} for $\mu_{\vec k}$. Inserting the second equation of Eq.~\eqref{eq:Theta_min} into Eq.~\eqref{eq:Full_small_Omega_k} yields Eq.~\eqref{eq:DefSmallOmega}.
     
     The magnon frequencies $\Omega_{\vec k}$ and the coefficients of the Bogoliubov transformation in Eq.~\eqref{eq:Bogoliubov_Parameters} are determined by solving the eigenvalue problem of 
     \begin{align}
         \begin{pmatrix}
            \omega_{\vec k}&\mu_{\vec k}^\ast\\
             -\mu_{\vec k}& - \omega_{-\vec k}
         \end{pmatrix}\begin{pmatrix}
             u_{\vec k}\\v_{\vec k}^\ast
         \end{pmatrix} =  \Omega_{\vec k}\begin{pmatrix}
             u_{\vec k}\\v_{\vec k}^\ast
         \end{pmatrix}\, .
     \end{align}
     
    \section{Calculation of the ground state \label{sec:appendixb}}
    \subsection{General interactions\label{appendixb1}} 
    Here, we derive Eqs.~\eqref{eq:Theta_min} and \eqref{eq:q0_min} as well as the direction of the cone axis direction $\vec n$, which will be represented by the angle variable $\vec n \cdot i\vec D_{\vec q_0}=\lvert\vec D_{\vec q_0}\rvert\cos\gamma$. We start from the classical energy given by Eq.~\eqref{eq:Classical_Energy} and define $\tilde E_{\text{cl}} = 2E_{\text{cl}}/(SN)$. This yields for the derivatives with respect to $(\vartheta, \vec q_0, \gamma)$
    \begin{align}
        \frac{\partial \tilde E_{\text{cl}}}{\partial \vartheta} &= -2 \sin(\vartheta) \left[\cos (\vartheta) A_{\vec q_0} - \muB \right] = 0\, ,\label{eq:E_derv_Theta}\\
        \frac{\partial \tilde E_{\text{cl}}}{\partial q_0^\alpha} &= \sin^2 (\vartheta)\partial_{q_0^\alpha}\left(J_{\vec q_0} + i D^{\vec{n}}_{\vec q_0} \right) = 0\, ,\label{eq:E_derv_q0}\\
        \begin{split}
            \frac{\partial \tilde E_{\text{cl}}}{\partial \gamma} &= i\sin^2 (\vartheta)\partial_{\gamma} D^{\vec{n}}_{\vec q_0} \\
            &=-\sin^2 (\vartheta)|\vec D_{\vec q_0}|\sin (\gamma ) = 0 \, .\label{eq:E_derv_Gamma}
        \end{split}
    \end{align}
    Equation~\eqref{eq:E_derv_Theta} is fulfilled for the solutions given in Eq.~\eqref{eq:Theta_min}. 
    As already mentioned in the main text, solving Eq.~\eqref{eq:E_derv_q0} requires knowing the specific form of the Fourier transforms, which will be performed for the square lattice in Appendix~\ref{appendixb2}. For the opening direction of the cone one obtains from Eq.~\eqref{eq:E_derv_Gamma} that $\vec n$ should be either parallel or antiparallel to $\vec D_{\vec q_0}$.
    
    Finding the minima from the stationary points satisfying Eqs.~\eqref{eq:E_derv_Theta}-\eqref{eq:E_derv_Gamma} may be performed by substituting them back in the energy expression, or by evaluating the Hessian
    \begin{align}
        H &= \begin{pmatrix}
    			H_{\vartheta}&0&0\\
    			0&H_{\gamma}&0\\
    			0&0&H_{\vec q_0}	
        \end{pmatrix}& H_{\alpha} &=\left. \frac{\partial^2  E_{\text{cl}}}{\partial^2 \alpha}\right|_{\text{ex}}  \, ,\label{eq:Hessian}
    \end{align}
    where $H_{\vec q_0}$ is the $d\times d$ matrix containing the second derivatives with respect to the components of $\vec q_0$, where $d$ is the dimension of the system. Note that the off-diagonal components of $H$ are generally non-zero, but they vanish in the stationary points. The second derivatives with respect to $\vartheta$ and $\gamma$ decouple and therefore need to be positive in the minimum.

    For $\vartheta$ this yields
    \begin{align}
        \frac{\partial^2 \tilde E_{\text{cl}}}{\partial \vartheta^2} &= 2 \left[\cos(\vartheta) \muB -\left(\cos^2(\vartheta) - \sin^2(\vartheta)\right)A_{\vec q_0} \right] > 0\, .
    \end{align}
    This implies that $\vartheta \in \{0,\pi\}$ will only be a minimum if $\muB >  A_{\vec q_0}$ or $\muB <  -A_{\vec q_0}$, and $\cos(\vartheta) = \muB/A_{\vec q_0}$ only for $A_{\vec q_0} > 0$ and $\mu_B g |B_{\mathrm n}| < |A_{\vec q_0}|$.
    
    The direction of $\vec n$ is determined by 
    \begin{align}
        \frac{\partial^2 \tilde E_{\text{cl}}}{\partial \gamma^2} &= -\sin^2 (\vartheta)|\vec D_{\vec q_0}|\cos (\gamma ) > 0\, ,
    \end{align}
    which together with Eq.~\eqref{eq:E_derv_Gamma} yields $\gamma = \pi$, and therefore 
    \begin{align}
        \vec n = -\frac{i\vec D_{\vec q_0}}{|\vec D_{\vec q_0}|} \,.
    \end{align}
    As mentioned in the main text, this condition holds for all wave vectors $\boldsymbol{q}_{0}$, and the direction of the external field will be rotated to agree with the $\vec{n}$ direction determined from this condition. To determine $\vec q_0$ via solving Eq.~\eqref{eq:E_derv_q0} and calculating the eigenvalues of $H_{\vec q_0}$, we need to specify the system and determine $J_{\vec q_0}$ and $\vec D_{\vec q_0}^{\vec{n}}$, which will be performed for the square lattice next. 
    
    \subsection{Square lattice\label{appendixb2}}
    \begin{figure}
    	\begin{center}
    			\includegraphics[width = \columnwidth]{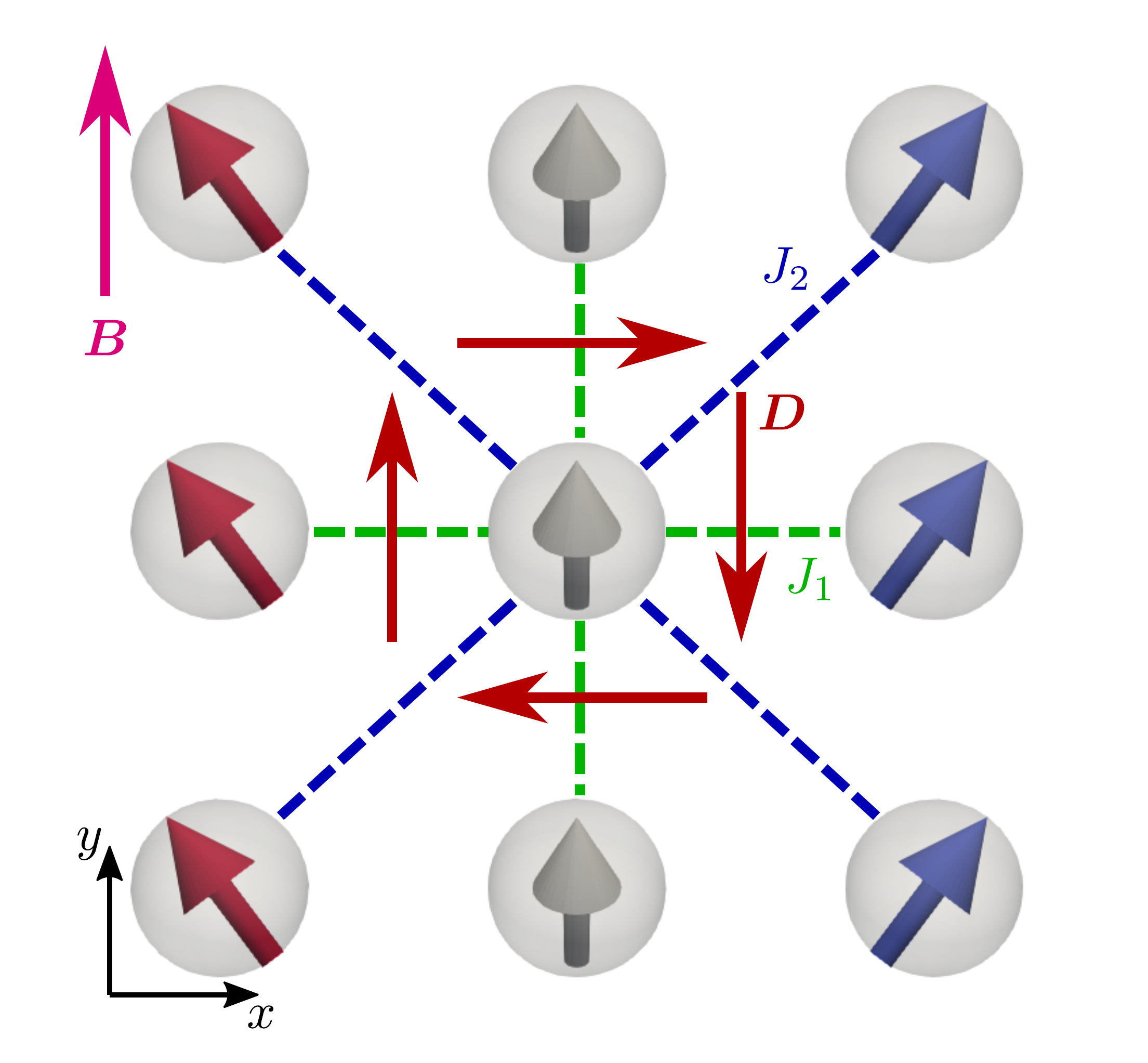}
    	\end{center}
    	\caption{Interactions in the square-lattice model. The interaction $J_1$ acts between the middle site and the neighboring sites (green lines, parallel to the $x$ ($y$) direction). The interaction $J_2$ acts between the middle site and the next-neighboring sites (blue, diagonal lines). The Dzyaloshinsky--Moriya vectors $\vec D$ (red arrows) between neighboring atoms are perpendicular to the connecting lines in the $C_{4\textrm{v}}$ symmetry class. The magnetic field $\vec{B}$ (purple arrow) is indicated in the upper left corner. Arrows inside the spheres illustrate the spin spiral.}
    	\label{fig:B1}
    \end{figure} 
    We regard the system described in Sec.~\ref{sec:3} and shown in Fig.~\ref{fig:B1}. The Fourier transform of the Heisenberg exchange reads
    \begin{align}
    \begin{split}
         J_{\vec k} &= \phantom{+}2J_1\left[\cos{(ak_x)} + \cos{(ak_y)}\right] +\\
        &\phantom{=}+  2J_2\left[\cos{(a(k_x+ k_y))} + \cos{(a(k_x - k_y))}\right]
    \end{split}  \\
    \begin{split}
        &= \phantom{+} 2J_1\left[\cos{(ak_x)} + \cos{(ak_y)}\right] +\\
        &\phantom{=} + 4 J_2\cos{(ak_x)}\cos{(ak_y)}\, .
    \end{split} 
    \end{align}
    For determining the directions of the Dzyaloshinsky--Moriya vectors, we assume a square-lattice magnet on a substrate with $C_{4\textrm{v}}$ symmetry. The substrate is necessary to break inversion symmetry between two spins, otherwise no DMI would be present. Following the rules for the DMI direction as listed by Moriya \cite{DMI_Moriya}, we conclude that $\vec D_{ij}$ points along the $x(y)$-direction if the vector connecting the two spins point along the $y(x)$-direction, as can be seen in Fig.~\ref{fig:B1}. Furthermore, using $\vec D_{ij} = - \vec D_{ji}$ we get 
    \begin{align}
        \vec D_{\vec k} &=  2i D \left(\begin{pmatrix}
            -\sin{(a k_y)}\\0\\0
        \end{pmatrix} + \begin{pmatrix}
            0\\ \sin{(a k_x)}\\0
        \end{pmatrix}\right)\, , \\
        D_{\vec q_0}^{\vec{n}} &= 2i|D|\sqrt{\sin^2{(aq_0^x)} + \sin^2{(a q_0^y)}}\, .
    \end{align}
    With the expressions for $J_{\vec k}$ and $\vec D_{\vec k}$, we can calculate the derivative of the classical energy Eq.~\eqref{eq:E_derv_q0} with respect to $q_0^x$,  
    \begin{align}
        \frac{\partial \tilde E_{\text{cl}}}{\partial q_0^x}  
        = -&2 a \sin^2 (\vartheta)\sin(a q_0^x)\Biggl [J_1   + 2 J_2 \cos(a q_0^y) + \nonumber\\&+\frac {|D| \cos(a q_0^x) }{\sqrt{\sin^2(a q_0^y) +\sin^2(a q_0^x)}}\Biggr ] = 0\,. \label{eq:Derv_E_q0_App}
    \end{align}
    One gets a similar equation for the derivative with respect to $q_0^y$ with the $x$ and $y$ components of $\vec q_0$ exchanged. 

    From these equations we are able to derive the SS configuration with different wave vectors discussed in Sec.~\ref{sec:3}, while the FP configuration stems from the first solution in Eq.~\eqref{eq:Theta_min}. In case of vanishing DMI, the stationary points are $q_{0}^{x}a\in\left\{0,\pi\right\}$ and $q_{0}^{y}a\in\left\{0,\pi\right\}$, and the minimum is given by the ferromagnetic state for $J_{2}/\lvert J_{1}\rvert <1/2$ and the row-wise antiferromagnetic state for $J_{2}/\lvert J_{1}\rvert >1/2$. For $J_{2}/\lvert J_{1}\rvert =1/2$, states with all values of $q_{0}^{x}$ are stationary for $q_{0}^{y}=0$, and they are also energetically degenerate. 

    For a finite DMI, the SS along $(1,0,0)$ is a stationary state if $\sin(aq_0^i) = 0$ with $i \in \{x,y\}$. For the other component of $\vec q_0$, which we will 
    call $q_0^{j}$, this would lead to
    \begin{align}
        J_1   + 2 r_i J_2   +\frac {|D| \cos(a q_0^{j}) }{|\sin(aq_0^{j})|} &= 0\, ,
    \end{align}
    where $r_i = \pm 1$ if $q_0^i = 0$ or $\pi$. Solving this equation yields $q_0^{j}$. 
    
    If we rewrite the system of equations \eqref{eq:Derv_E_q0_App} by subtracting the derivatives with respect to $q_{0}^{x}$ and $q_{0}^{y}$, we end up with 
    \begin{align}
  		2J_2\left[\cos(aq_0^x)- \cos(a q_0^y)\right] - \frac {|D| \left[\cos(aq_0^x) - \cos(a q_0^y)\right] }{\sqrt{\sin^2(a q_0^y) +\sin^2(a q_0^x)}} &= 0\, .
    \label{eq:Config_4_5_App}
    \end{align}
    This equation already yields the two other directions for the wave vector. For $\cos(aq_0^x)\neq \cos(a q_0^y)$, $\vec{q_{0}}$ is between the $(1,0,0)$ and $(1,1,0)$ directions, found as the solution of
    \begin{align}
        	\sqrt{\sin^2(a q_0^y) +\sin^2(a q_0^x)} &= \frac{|D|}{2 J_2}\, ,
    \end{align}
    together with 
    \begin{align}
        J_1 + 2J_2\left[\cos(aq_0^x) + \cos(a q_0^y)\right] &= 0\, ,
    \end{align}
    which is obtained when substituting the first equation into the sum over the derivatives along $q_{0}^{x}$ and $q_{0}^{y}$.
    The two equations yield that $0< -J_1/\left(2\sqrt{2}\right)  < J_2$ and $0< \lvert D\rvert/\left(2\sqrt{2}\right)  < J_2$ are required for this solution to be found. 

    Again regarding Eq.~\eqref{eq:Config_4_5_App} and the second possible solution of $q_0^x = \pm q_0^y = q_0$, which corresponds to a SS with wave vector along $(1,1,0)$, we find the sum of the derivatives with respect to $q_{0}^{x}$ and $q_{0}^{y}$ to yield 
    \begin{align}
        J_1 + 2J_2\cos{(aq_0)} + |D|\frac {\cos(aq_0) }{|\sin{(aq_0)}|} &= 0\, .
    \end{align}
    The case of $\sin(a q_0) = 0$ is excluded in all the above discussed cases as it would correspond to the case of the vanishing DMI. Determining $q_{0}$ from the above equation requires solving a quartic equation in $|\sin{(a q_0)}|$, which can be done analytically only in certain limits.  

    As mentioned in Appendix~\ref{appendixb1}, finding the global minimum can be performed by comparing the energies of the different stationary points. Calculating the eigenvalues of the Hessian can also be used to decide whether a certain stationary point is a minimum, but this is omitted here since the expressions are rather convoluted, it cannot be used to determine which of the local minima is the global minimum, and it is known in advance that one of the stationary points has to be the global minimum since the configuration space is compact.
    
\bibliography{bib}
 
\end{document}